\documentclass[journal,comsoc]{IEEEtran}
\usepackage{graphicx}
\usepackage{mathtools}
\usepackage{subcaption}
\usepackage{soul,xcolor}
\usepackage[normalem]{ulem}
\graphicspath{{figures/}} 

\newcommand{\boldmaxoper}{\mbox{\boldmath{Max:}}}
\newcommand{\boldvars}{\mbox{\boldmath{$(x,y,z)$}}}

\newcommand{\addturgay}[1]{\textcolor{black}{{ #1}}}

\begin{document}

\title{Heterogeneous GNN-RL Based Task Offloading for UAV-aided Smart Agriculture}
\author{Turgay Pamuklu$\dag$, \IEEEmembership{Member, IEEE}, \IEEEauthorblockN{Aisha Syed$\ddag$, W. Sean Kennedy$\ddag$, Melike Erol-Kantarci$\dag$, \IEEEmembership{Senior Member, IEEE}}

\IEEEauthorblockA{$\dag$\textit{School of Electrical Engineering and Computer Science,}
\textit{University of Ottawa}, Ottawa, Canada}

\IEEEauthorblockA{$\ddag$\textit{Nokia Bell Labs}\\
Emails:\{turgay.pamuklu, melike.erolkantarci\}@uottawa.ca, 
\{aisha.syed, sean.kennedy\}@nokia-bell-labs.com}
}

\maketitle

\makeatletter
\def\ps@IEEEtitlepagestyle{%
  \def\@oddfoot{\mycopyrightnotice}%
  \def\@oddhead{\hbox{}\@IEEEheaderstyle\leftmark\hfil\thepage}\relax
  \def\@evenhead{\@IEEEheaderstyle\thepage\hfil\leftmark\hbox{}}\relax
  \def\@evenfoot{}%
}

\def\mycopyrightnotice{
 \begin{minipage}{\textwidth}
 \centering \scriptsize
DOI: 10.1109/LNET.2023.3283936. © 20XX IEEE.  Personal use of this material is permitted.  Permission from IEEE must be obtained for all other uses, in any current or future media, including reprinting/republishing this material for advertising or promotional purposes, creating new collective works, for resale or redistribution to servers or lists, or reuse of any copyrighted component of this work in other works.
 \end{minipage}
}

\begin{abstract}
Having unmanned aerial vehicles (UAVs) with edge computing capability hover over smart farmlands supports Internet of Things (IoT) devices with low processing capacity and power to accomplish their deadline-sensitive tasks efficiently and economically. In this work, we propose a graph neural network-based reinforcement learning solution to optimize the task offloading from these IoT devices to the UAVs. We conduct evaluations to show that our approach reduces task deadline violations while also increasing the mission time of the UAVs by optimizing their battery usage. Moreover, the proposed solution has increased robustness to network topology changes and is able to adapt to extreme cases, such as the failure of a UAV.
\end{abstract}


\section{Introduction}
Cost-efficient Internet of Things (IoT) device deployment provides an economical and sustainable solution for smart agriculture. IoT devices can monitor events by handling alarms, and processing images or videos. Yet, the time-sensitive image or video processing and classification-based tasks need more advanced processing units than IoT devices can have onboard \cite{Nguyen2021}. Offloading these demanding tasks to a nearby unmanned aerial vehicle (UAV) is a promising approach to address this processing problem in rural settings. 

Several deep reinforcement learning (DRL) methods were proposed to solve this task offloading and resource allocation problem \cite{Li2020,Zhang2021,Nguyen2022,Pamuklu2022}. However, these approaches lack adaptability to uncertainties, such as the failure of a UAV. The generalization property of graph neural-network-based reinforcement learning (GNN-RL) methods can address these unpredictable cases \cite{Sun2021}. Recently, Guo et al. \cite{Guo2022} studied heterogonous graph convolutional networks for the transmission power control problem in radio access networks. Meanwhile, this heterogeneous model was not generic for all types of GNNs, and it was not addressing UAV-based networks. Li et al. have a recent study that proposed a GNN-RL solution for task offloading problems \cite{Li2022}. However, their solution is limited to a network model with only one UAV.

Unlike the previous studies, we solve the offloading problem of a multi-UAV-based environment using a generalized GNN-RL technique in this paper (Fig.~\ref{fig:concept1}). Our novel heterogeneous GNN-RL approach optimizes the offloading decisions to reduce the number of deadline violations and increase the mission time of the UAVs by optimizing their battery usage. Moreover, we train the neural network in an offline mode by using several cases before deploying it in a real-time environment. Due to the generalization property of the GNN \cite{Battaglia2018}, this offline-trained model gains robustness for unforeseen topology changes such as UAV failures. Therefore, with this model and approach, we achieve better performance than the traditional fully connected neural networks. To the best of our knowledge, this is the first work that provides a heterogeneous GNN-RL-based solution for a multi-UAV network.

The remainder of this letter is organized as follows. Section II and Section III describe the smart agriculture environment and the proposed GNN-RL solution, respectively. In Section IV, we present the results of our method and compare with three baselines. Finally, we conclude in Section V.

\section{System Model and Problem Definition}
\begin{figure*}[t]
  \begin{subfigure}{.30\textwidth}
  \includegraphics[width=1.0\textwidth]{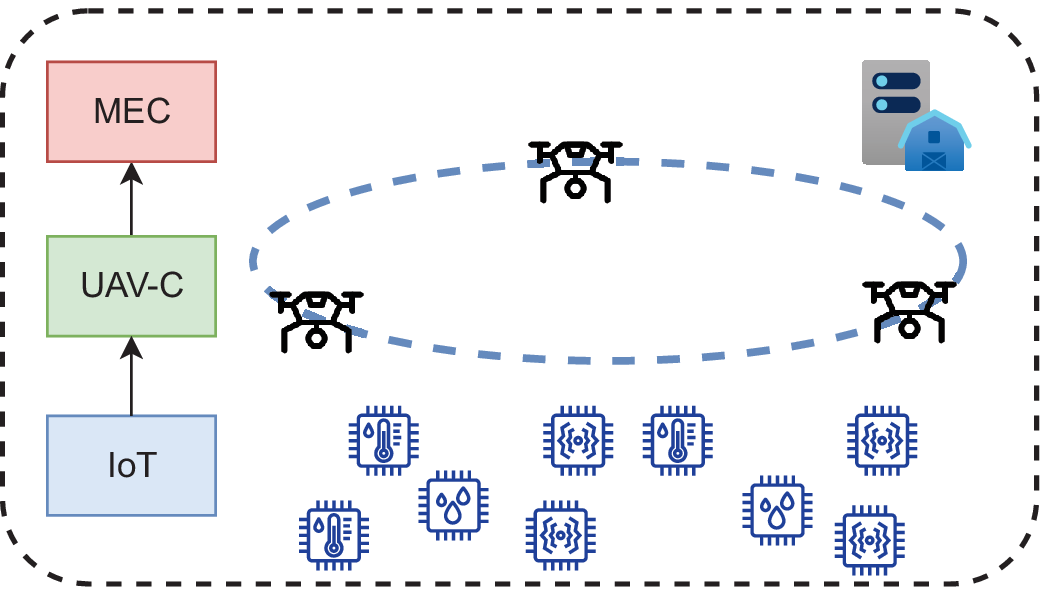}
  \caption{Smart agriculture environment.}\label{fig:concept1}
  \end{subfigure}
  \begin{subfigure}{.22\textwidth}
  \includegraphics[width=1.0\textwidth]{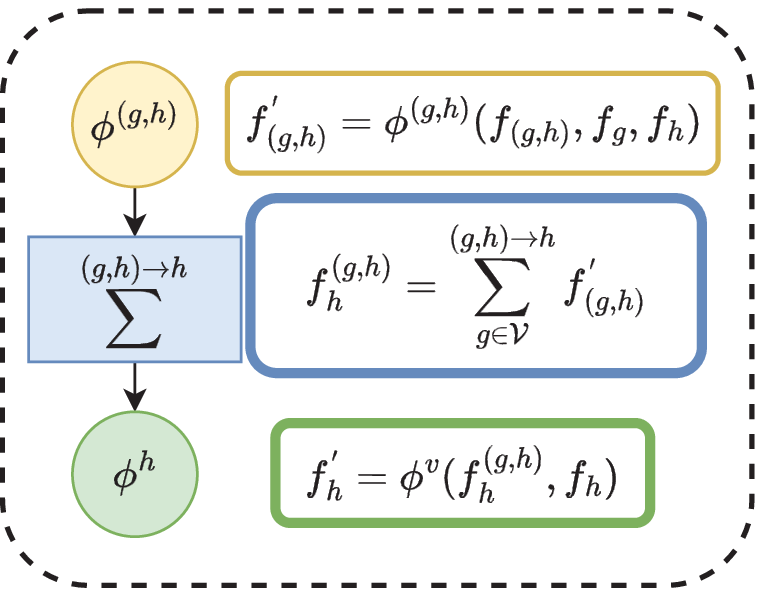}
  \caption{Graph network (GN) block.}\label{fig:concept2}
  \end{subfigure}
    \begin{subfigure}{.47\textwidth}
  \includegraphics[width=1.0\textwidth]{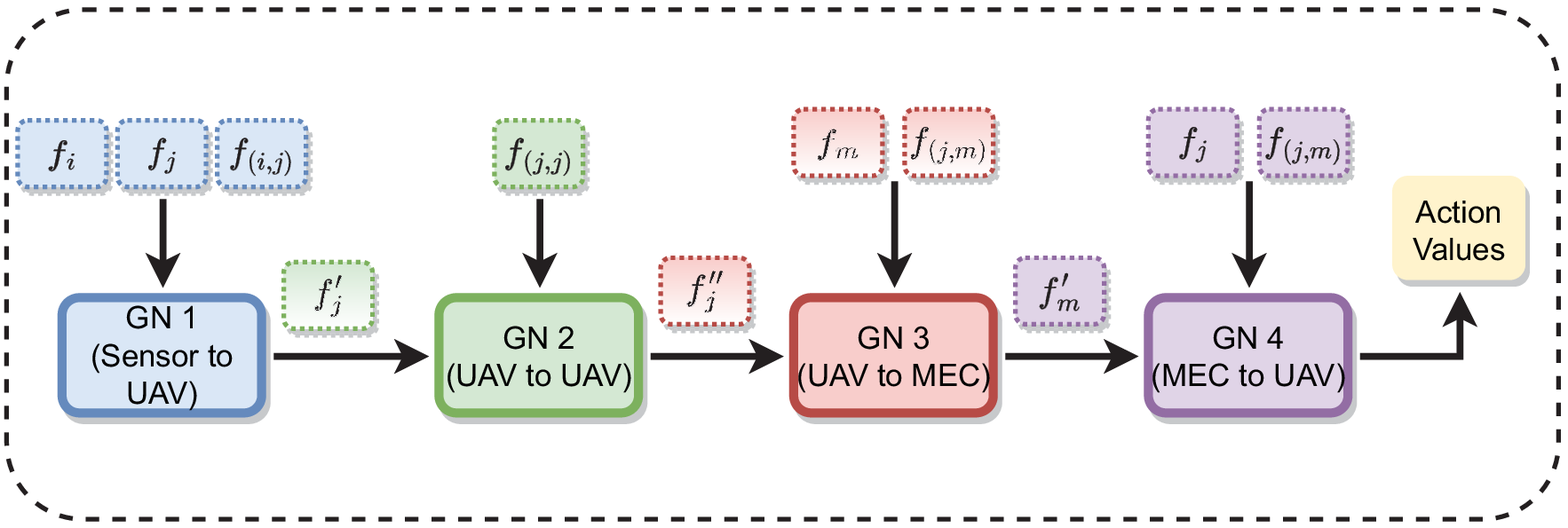}
  \caption{End-to-end GN block chain.}\label{fig:concept3}
  \end{subfigure}
    \caption{\label{fig:arch} Concept of three-tier heterogeneous GNN-RL based smart agriculture solution. IoT devices delegate their tasks to UAV-C devices. Then, UAV-Cs compute these tasks or offload them to another UAV-C or a more powerful terrestrial MEC device.}
\end{figure*}

\addturgay{Smart farming provides the capability to oversee the agriculture fields by utilizing sensors and cameras that transmit immediate updates on the status of their farmlands. With the help of image recognition technology, IoT devices can promptly detect alterations in the surroundings and potential hazards \cite{9838500}. As shown in Fig.~\ref{fig:concept1}, we consider $I$ IoT devices to observe farmland and create tasks in $T$ time intervals. These tasks are usually time-sensitive and must be completed before their deadlines. As a result, they are defined with a tuple $F_{i,t}=\{F_{i,t}^{P}, F_{i,t}^{L}, F_{i,t}^{D}\}$ where $F_{i,t}^{P}$, $F_{i,t}^{L}$ and $F_{i,t}^{D}$ are the packet size, processing load, and the deadline of a task ($i\in\mathcal{I}$) in time interval ($t\in\mathcal{T}$).}

To augment the limited computing capacity of the IoT devices, the network has $J$ computing UAVs (UAV-Cs) that travel on a given trajectory to collect tasks offloaded from the IoT devices. \addturgay{In other words, when an IoT device produces a task, it sends it to the closest UAV-C device. If the UAV-C's processing unit is on ($x_{j,t}=1$), it will handle the task locally. However, because of limited battery capacity ($\mathcal{B}{j}$), the UAV-C may pass the task to another nearby UAV-C with an active processing unit. Therefore, the first UAV-C can deactivate its task processing unit ($x_{j,t}=0$) in an underutilized time interval to conserve energy and thus maximize its mission time.}

In a farmland, UAV-Cs can have  Line-of-Sight (LoS) with IoT devices. Thus, probability of this communication profile can be calculated as \cite{Al-Hourani2014}:  
\begin{flalign}
\label{eq:losprob}
P^{LoS}(\theta_{i,j}) = & \frac{1}{1+a*exp(-b*(\theta_{i,j}-a))}&
\end{flalign}
Here, $a$ and $b$ are the macro-environment-related constant values, and $\theta_{i,j}$ is the angle between the locations of IoT $i$ and UAV-C $j$. Then the path loss is:
\begin{flalign}
\label{eq:pathloss}
L(d_{i,j}) = & 20log\left(\frac{4\Pi f_{c} d_{i,j}}{c}\right) + P^{LoS}\eta^{L} + (1-P^{LoS})\eta^{N}&
\end{flalign}
where $d_{i,j}$ is the distance between the IoT device $i$ and UAV-C $j$, $f_{c}$ is the carrier frequency, $c$ is the speed of light, and $\eta^{L}$ and $\eta^{N}$ are LoS and non-LoS related path losses, respectively. By using the path loss value, we can easily calculate the transmission data rate between IoT $i$ and UAV-C $j$ for a frequency division multiple access (FDMA) based channel with the Shannon-Hartley theorem:
\begin{flalign}
\label{eq:datarate}
r_{i,j} = & B_{i,j}*log_{2}\left(1+\frac{P^{tx}_{i}*L(d_{i,j})}{\sigma^{2}}\right) &
\end{flalign}
where $B_{i,j}$ is the allocated bandwidth for this channel, $P^{tx}_{i}$ is the transmission power, and $\sigma^{2}$ is the Gaussian noise power.
Eqs.~(\ref{eq:losprob}-\ref{eq:datarate}) are also used in the calculation of data rates between UAV-Cs, and from UAV-Cs to the terrestrial MEC. The only two differences are that the LoS probability is equal to one ($P^{LoS}=1$) for UAV-to-UAV communication and the transmission power of UAV-Cs ($P^{tx}_{j}$) is higher than that of the IoT devices.

We formulate the problem as a directed graph. This graph, $\mathcal{G}=(\mathcal{V}, \mathcal{E})$, represents the communication between the vertices ($\mathcal{V}=\{\mathcal{I}, \mathcal{J}, \mathit{m}\}$), where $(g,h)\in \mathcal{E}$ is a set of FDMA channels from the source vertex $g$ to the destination vertex $h$.  These channels have $\delta_{t}^{(g,h)}=\sum_{i\in\mathcal{I}}F_{i,t}^{P}*y_{i,t}^{(g,h)}*(1/r_{i,j})$ link delays where $y_{i,t}^{(g,h)}$ indicates that the task $F_{i,t}$ flows on edge $(g,h)$. Besides, the processing delays can be calculated by $\delta_{t}^{j}=\sum_{i\in\mathcal{I}}F_{i,t}^{L}*z_{i,t}^{j}*(1/C_{j})$ where $z_{i,t}^{j}$ indicates that the task $F_{i,t}$ is processed in UAV-C $j$ which has $C_{j}$ processing unit power. Thus, end-to-end delay for a task is calculated as:
\begin{flalign}
\label{eq:delay}
\delta_{i,t} = & \smashoperator{\sum\limits_{(g,h)\in \mathcal{E}}} \delta_{t}^{(g,h)} y_{i,t}^{(g,h)} + \smashoperator{\sum\limits_{j\in \{\mathcal{J}\cup{m}\}}} \delta_{t}^{j} z_{i,t}^{j}, \forall (i,t) \in (\mathcal{I},\mathcal{T})
\end{flalign}
A task is generated by an IoT device in a time interval $t$. Therefore, $(i,t)$ is a unique identification for a task, and the $\forall (i,t) \in (\mathcal{I},\mathcal{T}) $ definition covers all tasks in the network. However, Eq.~(\ref{eq:delay}) does not impose a flow constraint from the task sources (IoTs) to the destination (MEC). That constraint is maintained by Eq.~(\ref{eq:routing}), where $y_{i,t}^{(h,g)}$ is a binary decision variable that indicates task $F_{i,t}$ flows from vertex $h$ to vertex $g$.
\begin{multline}
\sum\limits_{h\in\mathcal{V}} y_{i,t}^{(g,h)} - \sum\limits_{h\in\mathcal{V}} y_{i,t}^{(h,g)}  =
\begin{cases} 
1, & \text{if } g\in \mathcal{I} \\
-1,& \text{if } g = \mathit{m} \\
0, & \text{otherwise}, 
\end{cases} \\ \forall (i,t) \in (\mathcal{I},\mathcal{T}), \forall g\in \mathcal{V}
\label{eq:routing}
\end{multline}

There are three cases according to the node types. If the node is a source node ($g\in\mathcal{I}$), the flow is only to the out direction; in that case, $y_{i,t}^{(g,h)}$ equals one, and $y_{i,t}^{(h,g)}$ equals zero. If the node is a destination node, the flow is only to the in direction ($y_{i,t}^{(g,h)}=0, y_{i,t}^{(h,g)}=1$). Otherwise, $y_{i,t}^{(g,h)}$ equals $y_{i,t}^{(h,g)}$, which means the flow passes through the intermediate nodes.

There are three main sources of energy consumption of a UAV-C. The first one is the propulsion energy ($\Upsilon^{P}_{j}$) that provides the movement of a UAV-C. Since there is a fixed flight route in our system, this value remains constant in our problem. The communication energy ($\Upsilon^{C}_{j}$), which is not considered in this study since it is significantly low compared to other energy sources \cite{Zeng2017}. Task processing is the third source of energy consumption. This study focuses on the optimization of data processing energy ($\Upsilon^{D}_{j}$). Eq.~(\ref{eq:encons}) shows the correlation of these energy sources and the remaining energy ($\Upsilon^{R}_{j}$) in the battery of UAV-C $j$, in which $\Upsilon^{B}_{j}$ is battery capacity. 
\begin{flalign}
\label{eq:encons}
\Upsilon^{R}_{j} = & \Upsilon^{B}_{j} - (\Upsilon^{P}_{j} + \Upsilon^{C}_{j}) * T - \sum\limits_{t\in\mathcal{T}}\Upsilon^{D}_{j}  * x_{j,t}&
\end{flalign}

The multi-objective function of the QoS-aware and mission time maximization problem is defined by Eq.~(\ref{eq:obj}), where $W$, $\Theta^{H}$, $\Theta^{D}$ are the weight and normalization factors, and $\mathbb{I}(.)$ is an indicator function which equals to one in case of a deadline violation. Eq.~(\ref{eq:demandconst}). Eq.~(\ref{eq:reachconst}) guarantee that task $F_{i,t}$ is processed by one and only one processing unit and that task flows on this processing unit, respectively. Lastly, Eq.~(\ref{eq:cpuconst}) activates the processing unit of UAV-C $j$ in case of task comes from an IoT to this UAV-C.
\begin{flalign}
\underset{\boldvars}{\boldmaxoper}&
\frac{W}{\Theta^{H}} * \min_{ j\in J} \Upsilon^{R}_{j} - \frac{(1-W)}{\Theta^{D}} * \sum\limits_{i\in\mathcal{I}}\sum\limits_{t\in\mathcal{T}} \mathbb{I}(\delta_{i,t} > F_{i}^{D}) \label{eq:obj}\\
\textbf{St: }&  \smashoperator{\sum\limits_{j\in\{\mathcal{J}\cup{m}\}}} z_{i,t}^{j} = 1, \qquad\qquad\qquad \forall (i,t) \in (\mathcal{I}, \mathcal{T}) \label{eq:demandconst} \\
&\smashoperator{\sum\limits_{j\in\{\mathcal{J}\cup{m}\}}} y_{i,t}^{(g,j)} * z_{i,t}^{j} = 1, \quad \forall (i,t) \in (\mathcal{I}, \mathcal{T}), \forall g \in \mathcal{V}  \label{eq:reachconst} \\
& \sum\limits_{i\in\mathcal{I}} z_{i,t}^{j} \leq M^{BIG} * x_{j,t}, \qquad\forall j \in \mathcal{J}, \forall t \in \mathcal{T} \label{eq:cpuconst} \\
&  Eq.~(\ref{eq:routing}) \notag
\end{flalign}

This maximization problem is a mixed-integer linear problem \cite{Joda2021}, and we proposed a GNN-RL-based solution, as detailed in the next section. 

\begin{figure}[t]
\centering
  \includegraphics[width=0.48\textwidth]{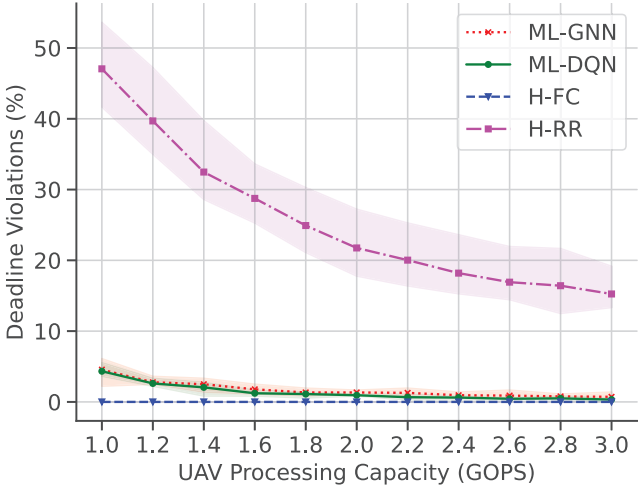}
  \caption{Impact of UAV-C processing capacity changes ($C_{i}$).}\label{fig:2a}
\end{figure}

\section{Heterogeneous GNN-RL Based Task Offloading Optimization}
We use the graph network (GN) blocks \cite{Battaglia2018} to structure the neural network model \footnote{In this section, time interval indexes have been removed from the equations to improve readability. It should be noted that the model is trained for each time interval.}. \addturgay{Fig.~\ref{fig:concept2} represents a GN block which starts with $\phi^{(g,h)}$ multilayer perceptron (MLP) that calculates the latent space $f_{(g,h)}^{'}$ of each $(g,h)$ edge.} The input of this perceptron is the concatenation of the features of this edge $f_{(g,h)}$, source vertex features $f_{g}$, and destination vertex features $f_{h}$. The second step $p^{(g,h)\to h}$ is a permutation invariant operation that calculates the summation of the edge latent spaces according to the destination vertex $h$. We concatenate the output of this step $f_{h}^{(g,h)}$ with the destination node features $f_{h}$. Then as a final step, we process that value with another MLP operation ($\phi^{h}$). This model is suitable for parallel processing of the edges and vertices as a mini-batch which provides remarkable scalability for larger networks. 

Our smart agriculture network has different types of vertices; thus, a standard homogenous GNN model cannot address the optimization problem of this network. Therefore, we model a novel heterogeneous GNN for each relation between these vertices by using GN chains. \addturgay{Fig.~\ref{fig:concept3} shows this chain in which the first GN block in this chain provides the relational inductive bias between the IoTs and UAV-Cs.} Here, the source vertex features $\mathbf{f_{i}}=[ \mathbf{F_{i}^{P}}\mathbin\Vert \mathbf{F_{i}^{L}} \mathbin\Vert \mathbf{F_{i}^{D}}]$ are the task tuples of the IoT nodes $i \in \mathcal{I}$; the destination vertex features $\mathbf{f_{j}}=[\mathbf{\Upsilon^{R}_{j}}\mathbin\Vert \mathbf{\delta^{j}}\mathbin\Vert \mathbf{C_{j}} ]$ are a concatenation of the remaining energy, processing delay, and processing capacity vectors of UAV-Cs. The edge features $\mathbf{f_{(i,j)}}=[\mathbf{\delta^{(i,j)}}\mathbin\Vert \mathbf{\mathcal{C}^{X}_{j}}\mathbin\Vert \mathbf{\mathcal{C}^{Y}_{j}} ]$ are the link delays and cartesian coordinates of the UAV-Cs connected to these links.  
The second GN block provides an induction between UAV-Cs. This block receives the UAV-C latent features ($\mathbf{f'_{j}}$) from the previous GN block by using the ``graph-to-graph" property of the GN block \cite{Battaglia2018}. Then, the model integrates these features as source and destination vertice features and includes the link features between these UAV-Cs as edge features $\mathbf{f_{(j1,j2)}}= [\mathbf{\delta^{(j1,j2)}}\mathbin\Vert \mathbf{\mathcal{C}^{X}_{j2}}\mathbin\Vert \mathbf{\mathcal{C}^{Y}_{j2}}]$.

\begin{figure}[t]
\centering
  \includegraphics[width=0.48\textwidth]{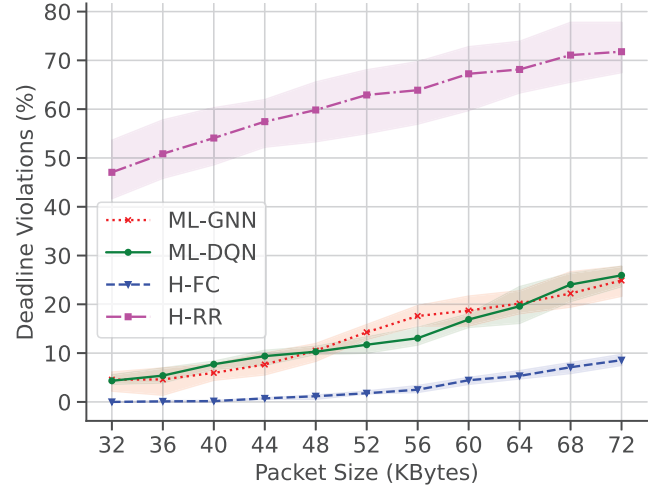}
  \caption{Impact of task packet size changes  ($F_{i}^{P}$).}\label{fig:2b}
\end{figure}

The third GN block computes the relation between UAV-Cs and the MEC vertex. The block receives the output of the second block to use it as the source vertices (UAV-Cs) features ($\mathbf{f''_{j}}$) and includes the MEC node delays\footnote{In the implementation phase, we extend the size of the MEC node features to the largest ones, and concatenate with zeros for a GPU-based implementation related performance improvement. This approach speeds up the learning process.} $\mathbf{f_{m}}=[ \mathbf{\delta^{x}}\mathbin\Vert \mathbf{0}\mathbin\Vert \mathbf{0} ]$ and the link features between the UAV-Cs and MEC ($\mathbf{f_{(j,m)}}=[\mathbf{\delta^{(j,m)}}\mathbin\Vert \mathbf{\mathcal{C}^{X}_{j}}\mathbin\Vert \mathbf{\mathcal{C}^{Y}_{j}} ]$). Next, we use a UAV-C node classification approach to provide a robust machine learning (ML) model due to UAV-C failures. Thus, we include the fourth GN block, which uses the output of the third GN block as the source node feature and the UAV-C features as the destination node features. The output of this GN block is the state action values of processing unit activation decisions ($x_{j}$).

The RL framework uses the above values to calculate the loss function and serves as a traditional DQN framework to optimize these decisions according to the objective function. Therefore, the objective function Eq.~(\ref{eq:obj}) is converted to the immediate reward function Eq.~(\ref{eq:reward}) for the training process, where $\Upsilon^{S}_{t}$ is the severity level of the UAV-C that has the lowest remaining energy. This value is calculated by Eq.~(\ref{eq:rew2}), where the severity condition occurs when the UAV-C that has the lowest remaining energy ($\min\limits_{j\in J}(\Upsilon^{R}_{j})$) has an active processing unit ($x_{j}=1$). In this case, negative feedback is provided, which is quadratically increasing with the difference between the remaining energy and the battery capacity. Otherwise, positive feedback is returned to promote the decision ($x_{j}=0$).  
\begin{flalign}
\label{eq:reward}
\mathcal{R}_{t} = & \frac{W}{\Theta^{H}} * \Upsilon^{S}_{t} - \frac{(1-W)}{\Theta^{D}} * \mathbb{I}(\delta_{i,t} > F_{i}^{D}) \\
\label{eq:rew2}
& \Upsilon^{S}_{t} =
\begin{cases} 
        \frac{\left(\Upsilon^{R}_{j}-\mathcal{B}_{j}\right)^{2}}{\left(\mathcal{B}_{j}\right)^{2}}, & \text{if } \sum\limits_{j \in \mathcal{J}} x_{j} * \arg \min\limits_{j\in J}(\Upsilon^{R}_{j}) = j, \\
        1, & \text{otherwise.}
\end{cases}
\end{flalign}

We follow the learning rate decay technique for the optimizer and decayed epsilon-greedy for choosing the actions. Also, a target network is added to improve the model stability.

\begin{figure}[t]
\centering
  \includegraphics[width=0.48\textwidth]{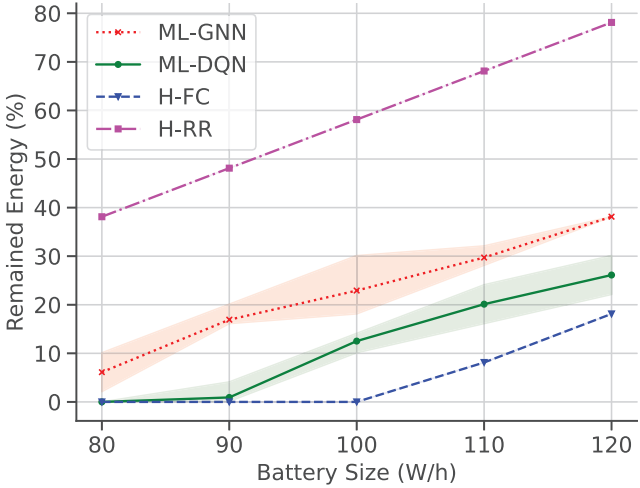}
  \caption{Impact of task processing load ($F_{i}^{L}$).}\label{fig:2c}
\end{figure}

\section{Computational Experiments}
The simulation environment consists of $I=24$ IoT devices, $J=8$ UAV-Cs, one MEC and $T=18$ time intervals. UAV-Cs follow a grid-based route at a fixed altitude ($5$ m), and they have the same size of batteries ($\mathcal{B}_{j}=100$ W/h) and propulsion energy consumptions ($\Upsilon^{P}_{j}=20$ W/h). The physical layer propagation parameters ($a=4.88$, $b=0.43$, $f_{c}=2$ GHz) are chosen for a suburban environment \cite{Bor-Yaliniz2016}. The IoT devices generate three types of tasks ($F_{i,t}^{D}=[6, 10, 15]$ s) with a Poisson arrival distribution (0.5 s mean arrival rate), and the processing times of these tasks are exponentially distributed with a mean ($F_{i,t}^{L}=[0.6, 1, 1.5]$ s). The packet size of these three different task types are equal ($F_{i,t}^{P}=32$ KBytes).

\begin{figure}[t]
\centering
  \includegraphics[width=0.48\textwidth]{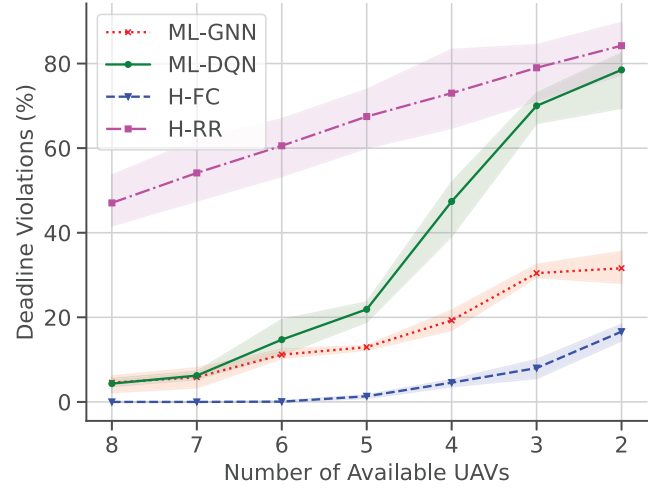}
  \caption{Impact of UAV-C Failures.}\label{fig:2d}
\end{figure}

The Heterogenous GNN-RL model is implemented with Pytorch library \cite{Paszke2019}, and Compute Canada GPU clusters \cite{Canada2022} are used to train the model. The neural network model parameters are chosen as mini-batch with size $128$, target network updates per $50$ steps, and the optimizer used is Adam with a decay learning technique. The number of episodes equals to $50,000$ with a decayed epsilon-greedy exploration and $0.95$ gamma value.

We use three baselines for evaluating our model. First is \textit{heuristic full-time computing} or ``H-FC" in which each UAV-C's processing unit is continuously active during the simulation. Therefore, each UAV-C is ready to serve the IoT devices. Obviously, this heuristic yields a lower-bound for the number of deadline violations. The second baseline is \textit{heuristic round-robin} or ``H-RR" in which only one UAV-C's processing unit is active in a time interval. We choose the active unit with a round-robin scheme. Hence, this heuristic is an upper bound for the mission time. Lastly, we use a \textit{fully-connected ML approach} or ``ML-DQN" as the third baseline.

We trained ML-GNN and ML-DQN in offline mode with several task sets, and then we used these ML models in inference mode to get the following experimental results. This approach helps to train the models with edge cases. \addturgay{The implementation takes a centralized approach in which the MEC device is selected as the deployment location for the neural network. We also presume that control channels exist between the MEC device and UAV-Cs to allow for the sharing of control information between these devices.}

Fig.~\ref{fig:2a} shows the impact of processing capacity changes ($C_{i}$) in the baselines and in our proposed ML-GNN model in terms of deadline violations. While the trends for all methods are decrementation with increasing capacity, ``H-RR" lacks the capability of reducing these violations significantly even under decent processing capacity.  ML methods provide close performance results to each other and they are close to the lower bound of ``H-RR." Fig.~\ref{fig:2b} represents the impact of packet size, which increases the delay between the wireless channels of different entities of the network. ML solutions achieve similar performances that are near the lower bound provided by H-RR.   

A significant performance change in ML-GNN can be seen for mission time improvement. Fig.~\ref{fig:2c} shows the remaining energy of the UAV-C with lowest energy. It is clearly seen that ML-GNN outperforms ML-DQN even with larger battery sizes $\mathcal{B}_{j}=100$ W/h. In inference mode, we also reduce the number of available UAV-Cs in the environment. In that case, ML-GNN performs much better than ML-DQN for balancing the task loads between the other UAV-Cs (Fig.~\ref{fig:2d}). Therefore, the proposed model has higher robustness for topology changes and UAV-C failures.

\begin{figure}[t]
\centering
  \includegraphics[width=0.48\textwidth]{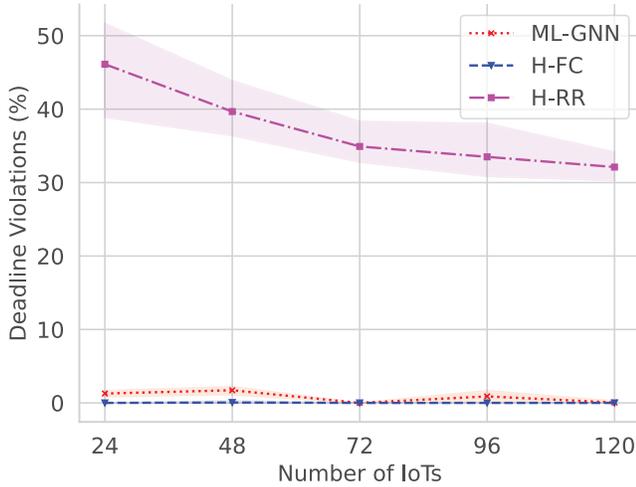}
  \caption{\addturgay{Scalability Analysis.}}\label{fig:3}
\end{figure}

\addturgay{Furthermore, in addition to the UAV-C failure test case, we have evaluated the scalability of the proposed model as the number of IoT devices increases. As previously stated, we trained the proposed neural network using 24 IoT devices with predetermined processing times ($F_{i,t}^{L}$) and packet sizes ($F_{i,t}^{P}$). We then utilized this trained neural network in inference mode with an increasing number of IoT devices. To maintain similar processing and transmission delays, we adjusted the processing times and packet sizes proportionally to the number of IoT devices. The outcomes demonstrate that our proposed neural network can be expanded to accommodate larger networks without any degradation in performance, as shown in Fig.~\ref{fig:3}.}

\section{Conclusion}
This study introduces a novel approach with a GNN-RL based machine-learning model for solving the IoT-UAV-based smart farmland task offloading problem. The approach shows superior performance at addressing the tradeoff between reducing the number of deadline violations and increasing the mission time objectives. Also, due to the generalization property of the graph structure, the GNN-RL model has better adaptability for topology changes that happen during the inference mode. Moreover, the model can be easily generalized to other wireless network problems due to its heterogeneous architecture.

\addturgay{In our future work, we plan to combine the multi-task learning concept with this generalized GNN-RL architecture. Furthermore, we plan to transform the current centralized implementation into a multi-agent reinforcement learning implementation utilizing a partial observation approach.}

\section*{Acknowledgment}
This  work  is  supported  by  MITACS Canada Accelerate program under collaboration with Nokia Bell Labs.

\bibliography{main}

\begin{thebibliography}{10}

\bibitem{Nguyen2021}
A.~C. Nguyen, T.~Pamuklu, A.~Syed, W.~S. Kennedy, and M.~Erol-Kantarci,
  ``{Reinforcement Learning-Based Deadline and Battery-Aware Offloading in
  Smart Farm IoT-UAV Networks},'' in {\em ICC - International Conference on
  Communications}, IEEE, 2022.

\bibitem{Li2020}
K.~Li, Y.~Emami, W.~Ni, E.~Tovar, and Z.~Han, ``{Onboard Deep Deterministic
  Policy Gradients for Online Flight Resource Allocation of UAVs},'' {\em IEEE
  Networking Letters}, vol.~2, pp.~106--110, sep 2020.

\bibitem{Zhang2021}
L.~Zhang, Z.-Y. Zhang, L.~Min, C.~Tang, H.-Y. Zhang, Y.-H. Wang, and P.~Cai,
  ``{Task Offloading and Trajectory Control for UAV-Assisted Mobile Edge
  Computing Using Deep Reinforcement Learning},'' {\em IEEE Access}, vol.~9,
  pp.~53708--53719, 2021.

\bibitem{Nguyen2022}
A.~C. Nguyen, T.~Pamuklu, A.~Syed, W.~S. Kennedy, and M.~Erol-Kantarci, ``{Deep
  Reinforcement Learning for Task Offloading in UAV-Aided Smart Farm
  Networks},'' in {\em IEEE FNWF}, 2022.

\bibitem{Pamuklu2022}
T.~Pamuklu, A.~C. Nguyen, A.~Syed, W.~S. Kennedy, and M.~Erol-Kantarci,
  ``{IoT-Aerial Base Station Task Offloading with Risk-Sensitive Reinforcement
  Learning for Smart Agriculture},'' {\em IEEE Transactions on Green
  Communications and Networking}, pp.~1--1, 2022.

\bibitem{Sun2021}
P.~Sun, J.~Lan, J.~Li, Z.~Guo, and Y.~Hu, ``{Combining Deep Reinforcement
  Learning With Graph Neural Networks for Optimal VNF Placement},'' {\em IEEE
  Communications Letters}, vol.~25, pp.~176--180, jan 2021.

\bibitem{Guo2022}
J.~Guo and C.~Yang, ``{Learning Power Allocation for Multi-Cell-Multi-User
  Systems with Heterogeneous Graph Neural Networks},'' {\em IEEE Transactions
  on Wireless Communications}, vol.~21, no.~2, pp.~884--897, 2022.

\bibitem{Li2022}
K.~Li, W.~Ni, X.~Yuan, A.~Noor, and A.~Jamalipour, ``{Deep Graph-based
  Reinforcement Learning for Joint Cruise Control and Task Offloading for
  Aerial Edge Internet-of-Things (EdgeIoT)},'' {\em IEEE Internet of Things
  Journal}, pp.~1--1, 2022.

\bibitem{Battaglia2018}
P.~W. Battaglia, J.~B. Hamrick, V.~Bapst, A.~Sanchez-Gonzalez, V.~Zambaldi,
  M.~Malinowski, A.~Tacchetti, D.~Raposo, A.~Santoro, R.~Faulkner, C.~Gulcehre,
  F.~Song, A.~Ballard, J.~Gilmer, G.~Dahl, A.~Vaswani, K.~Allen, C.~Nash,
  V.~Langston, C.~Dyer, N.~Heess, D.~Wierstra, P.~Kohli, M.~Botvinick,
  O.~Vinyals, Y.~Li, and R.~Pascanu, ``{Relational inductive biases, deep
  learning, and graph networks},'' pp.~1--40, 2018.

\bibitem{9838500}
A.~C. Nguyen, T.~Pamuklu, A.~Syed, W.~S. Kennedy, and M.~Erol-Kantarci, ``To
  risk or not to risk: Learning with risk quantification for iot task
  offloading in uavs,'' in {\em ICC 2023 - IEEE International Conference on
  Communications}, pp.~189--194, 2023.

\bibitem{Al-Hourani2014}
A.~Al-Hourani, S.~Kandeepan, and S.~Lardner, ``{Optimal LAP Altitude for
  Maximum Coverage},'' {\em IEEE Wireless Communications Letters}, vol.~3,
  pp.~569--572, dec 2014.

\bibitem{Zeng2017}
Y.~Zeng and R.~Zhang, ``{Energy-Efficient UAV Communication With Trajectory
  Optimization},'' {\em IEEE Transactions on Wireless Communications}, vol.~16,
  pp.~3747--3760, jun 2017.

\bibitem{Joda2021}
R.~Joda, M.~Elsayed, H.~Abou-Zeid, R.~Atawia, A.~B. Sediq, G.~Boudreau, and
  M.~Erol-Kantarci, ``{Carrier Aggregation With Optimized UE Power Consumption
  in 5G},'' {\em IEEE Networking Letters}, vol.~3, pp.~61--65, jun 2021.

\bibitem{Bor-Yaliniz2016}
R.~I. Bor-Yaliniz, A.~El-Keyi, and H.~Yanikomeroglu, ``{Efficient 3-D placement
  of an aerial base station in next generation cellular networks},'' in {\em
  2016 IEEE International Conference on Communications (ICC)}, pp.~1--5, IEEE,
  may 2016.

\bibitem{Paszke2019}
A.~Paszke, A.~Lerer, T.~Killeen, L.~Antiga, E.~Yang, S.~Gross, J.~Bradbury,
  F.~Massa, and B.~Steiner, ``{Pytorch: An imperative style, high-performance
  deep learning library},'' in {\em Advances in neural information processing
  systems}, no.~NeurIPS, p.~32, 2019.

\bibitem{Canada2022}
{Digital Research Alliance of Canada}, ``{Compute Canada},'' 2022.
\newblock Available at https://alliancecan.ca/, Accessed Date: 24-10-2022.

\end{thebibliography}
\end{document}